\documentstyle[11pt,moriond,epsfig]{article}

\def\be{\begin{equation}}
\def\ee{\end{equation}}
\def\bea{\begin{eqnarray}}
\def\eea{\end{eqnarray}}

\newcommand{\thomp}{{4
    \pi^2 \over 45}\, c \,\sigma_t}  
\begin{document}
\vspace*{4cm}
\title{AVATARS OF A MATTER--ANTIMATTER UNIVERSE
\footnote{Based on a talk at the January 1998 Moriond Meeting
at Les Arcs, France.}}

\author{ALVARO DE R\'UJULA}

\address{TH Division, CERN, CH-1211 Geneva}
\maketitle\abstracts{An elegantly symmetric Universe,
 consisting of large islands
of matter and antimatter, is by no means obviously
out of the question. I review the observations that lead to the usual 
{\it prejudice} that the Universe contains only matter. 
I discuss recent work  \cite{CDGl} inferring that this prejudice can be converted
into an inescapable conclusion. I argue that our theoretical conviction should
not discourage direct searches \cite{AMS} for antimatter
in cosmic rays.}


  \normalsize\baselineskip=15pt
  \setcounter{footnote}{0}
  \renewcommand{\thefootnote}{\alph{footnote}}

\section{Evidence against cosmic antimatter}

The Universe contains lots of light (some 400 microwave background 
photons per cc), a little matter (a few baryons and electrons for every billion
photons) and practically no antimatter, at least in our neighbourhood.
This is a surprisingly unbalanced recipe. I discuss how, 
on the basis of empirical observations and conventional
physics, and 
on very general theoretical grounds, one can {\it prove}
that our Universe contains no significant amounts of 
antimatter \cite{CDGl}.

The possibility that there be matter and antimatter
``islands'' in the Universe has received occasional attention
\mbox{\cite{Counterex}}. Early attempts to construct such a theory, notably by
Omn\`es \mbox{\cite{Omn}},
were not successful. They faced the impossible
and self-imposed task of separating the
ingredients from their mixture. For domains of
matter and antimatter to be present, baryogenesis and
``antibaryogenesis'' must have occurred in large separate domains.

In 1976 Steigman thoroughly reviewed the theory and observations of
cosmic antimatter \mbox{\cite{Steigman}}. This work is still very
much up to date, particularly in its recounting of observational
constraints; much of the discussion in this section relies on it.

\subsection{Antimatter about our planet}

Various balloon- and satellite-borne detectors have observed cosmic-ray positrons and antiprotons. Their flux is
compatible with the expectation for the secondary
products of conventional (matter)
cosmic rays impinging on interstellar matter (gas and dust). The
$\bar{p}/p$ ratio is expected to diminish precipitously below a
kinetic energy of a few GeV: at the high energy required to
produce these secondaries, the production of a slow $\bar{p}$ is
unlikely. The 1981 observation \mbox{\cite{BSP}} of a large $\bar{p}$
excess at $E_{_{\rm KIN}}\sim\! 0.1$ GeV created a stir:
it could be fashionably interpreted as the result of
galactic-halo dark-matter neutralino--neutralino annihilation. 
Subsequent observations \mbox{\cite{ASSS,Golden}} 
brought the $\bar{p}$ flux back to the standard
expectations. There is also a hint --in no way significant-- of
a low-energy positron excess \mbox{\cite{Review1,Golden}}.

The cosmic-ray flux of many different nuclei is well measured in a
domain of kinetic energy (per nucleon) extending from a few MeV to
circa 1 TeV. But for a small fraction of anti-deuterons, one does not expect an observable flux of antinuclei,
for the energy required to make these fragile objects in
matter--matter collisions is far in excess of their binding energy. No
convincing observation of $Z>2$ antinuclei has been
 reported. It is
often
emphasized that the observation of a single antinucleus would
be decisive evidence for an antimatter component of the Universe:
it is likely that $\overline{{\rm He}}$ would be the result of
primordial antinucleosynthesis;  $\overline{{\rm C}}$ would
presumably originate in an antistar.

\subsection{Antimatter in our galaxy}

The picture of Armstrong's footprint on the lunar surface is the most
convincing evidence that the astronaut and the Moon were made of the
same stuff.  The planets, asteroids and comets
of the solar system are also of uniform
composition. The solar wind (mainly protons) would otherwise shine in
observable gamma rays as it impinges on their surfaces or
atmospheres.

Observations of the 0.511 MeV $e^+ e^-$ annihilation line in the direction 
of the galactic center,
particularly by the OSSE instrument on board of the
Compton GRO satellite \cite{Skibo}, show an interesting
distribution of annihilating positrons. Their origin can be quite
unsurprising:  $\beta^+$ decay products of
elements synthesized in supernovae, novae, and Wolf-Rayet stars
\cite{Dermer}.

Of the various constraints on a possible contamination of antibaryons
in our galaxy, the most stringent pertains to observations of hydrogen
in ``clouds''. Gamma rays from their directions are
observed; they are compatible with $\pi^0$ decay, the pions being
secondary products of collisions of ordinary cosmic rays with the
hydrogen in the clouds. The non-observation of a $\gamma$ excess
implies that the antibaryon contamination in these media cannot
exceed one part in $10^{15}$, an astonishingly stringent result.

Galaxies are supposed to have undergone a phase of
recollapse onto themselves, after they lagged behind the general
Hubble expansion to become objects of fixed size, at a redshift of a
few. This recollapse is reckoned to mix and virialize the galactic
material, and to re-ionize its ordinary matter.
If this process could occur in a galaxy containing matter and antimatter
(a possibility to be doubted anon) it would annihilate the minority
ingredient, or blow the galaxy apart.
Nonetheless, the search for ordinary or neutron antistars is of
interest, for they may not ``belong'' to our galaxy, but be intruders
from afar. These objects would accrete interstellar gas and shine
$\gamma$ rays. At the time of Steigman's review, the point-like and
diffuse limits were: no single antistar in our 30 parsec
neighbourhood, no more than one antistar for every 10$^4$ ordinary ones.
I do not know whether these limits have been subsequently updated.

\subsection{Antimatter beyond our galaxy}

The
``photonic'' astronomer cannot
 determine whether or not another galaxy is
made of matter or of antimatter. 
But  galaxies in collision are often observed: 
the Antennae pair
NGC4038(9) is a gorgeous example. An encounter involving
a galaxy and an antigalaxy would be spectacular. No such
event has been reported.

The largest objects on whose ``purity'' we have information are
clusters of galaxies. Some of them are sufficiently dense and active
to sustain an intergalactic hot plasma in their central parts, at a
temperature of order 10 keV. The failure to observe a $\gamma$-ray
excess atop the thermal spectral tail implies a purity at a level of
a few   parts per
million.
Clusters of galaxies (of a typical size O(20) Mpc, or $6\times
10^7$   light years) are the largest objects 
empirically known to have a uniform
composition.

In 1971 Stecker {\it et al.} \mbox{\cite{SMB}} studied a
matter--antimatter Universe in which, somehow, a {\it fixed}
(time-independent)
fraction of the average baryon (or antibaryon) density is
continuously annihilating. They could choose that fraction
so that the (redshifted) photons from $\bar{p} p$ annihilation would
reproduce the (then) observed shoulder or hump 
at $\sim\! 2$ MeV in the cosmic
diffuse $\gamma$-ray spectrum (CDG), see Fig. 1. The ingredients of their
cosmology are by now outdated, and the presence of the 
hump, as we shall see, is debated \mbox{\cite{Newspectrum}}.

\begin{figure}
\epsfysize=4.0in
\centerline{\epsffile{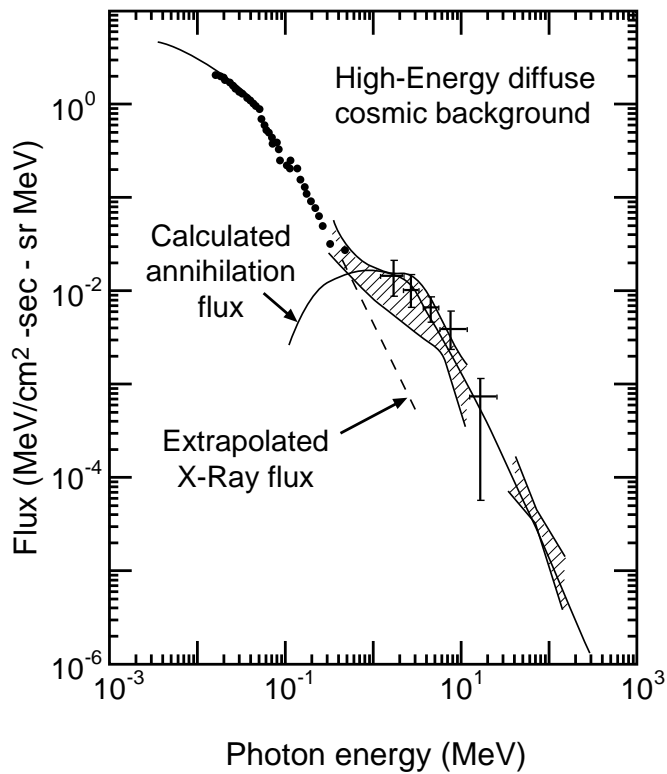}}
\caption{Energy flux of the diffuse $\gamma$-ray spectrum.}
\label{fig:sizes}
\end{figure}



\section{Conclusions at this early stage}

Define a {\it domain\/} to be a region of the Universe  whose typical
dimension today is $d_0$, with $d_0\ge 20$~Mpc. The radial extension of
the
visible Universe is a few thousand Mpc;  it contains as
many
as 30 million domains. Given the   observed abundance of galaxies,
each
domain would in turn contain, on   average, at least a few thousand
galaxies or antigalaxies.

We have seen that there are also empirical arguments whereby clusters of galaxies
(regions similar to or smaller than the domains we have defined) ought
to be made of either matter or antimatter. Our own domain, by
definition, is made of matter. It is a common but considerable
extrapolation to conclude from this single example that the rest of
the tens of millions of visible domains are also made of {\it
matter}.

Imagine a patchwork Universe made of regions of matter and
antimatter. Suppose that each
 region is  currently larger than 20 Mpc and contains either
galaxies
or antigalaxies.
At this point of my talk, it is {\bf NOT} observationally excluded that some domains in the
Universe (say half of them) be made of antimatter. Thus, it behooves
one:
\begin{itemize}
\item{to try and improve the 20 Mpc  limit on the domain size $d_0$,
by reexamining the constraints imposed by current 
observations \cite{CDGl};}
\item{to try and improve the $d_0>20$ Mpc limit by undertaking novel observations
 \cite{AMS}.}
\end{itemize}

\section{Extant views on  cosmo- and baryogenesis}

At a very early age, it is argued, the Universe ``inflated''
superluminally, with a scale factor
$R(t)\propto t^n,\;  n>1$; or $R(t)\propto
\exp[H\, t]$. Inflation can tell why the cosmic background radiation
(CBR) is so uniform, how the cosmic structures may have evolved from
inevitable quantum fluctuations, why the density of the Universe at
its ripe old age (in the natural Plank-time units) is so close to
critical, how its enormous entropy could have evolved from a primeval
state of extreme simplicity ... No alternative scheme competes with inflation
in its explanatory power.

In the conventional picture, baryon--number--violating interactions are
prevalent in the very early Universe, so that the concept of an
initial
baryon number in not meaningful.
As Sakharov prophesied, the interplay of CP- and baryon-number-violating interactions, in a period far from thermal equilibrium,
could
have led to a universal baryon asymmetry; that is, to
 baryon and antibaryon densities satisfying
$n_B-n_{\bar B}>0$ {\it everywhere.}  In this scenario, there
would
be today no cosmologically significant amounts of
 antimatter in the visible Universe.

It goes without saying that in order to imagine an
unconventional  Universe consisting of matter and antimatter domains,
 several observational and theoretical constraints must be satisfied:
sufficient inflation must have taken place to reproduce the
successes
of standard inflationary scenarios;
the standard picture of primordial nucleosynthesis must
survive
unscathed;
a mechanism must be present to generate the 
inhomogeneities seen in the CBR and 
those argued to be the seeds of subsequent structure
formation;
the CBR must be close enough to thermal and uniform
to conform to observations; the surviving flux of $\gamma$-rays from 
matter--antimatter annihilations at domain
boundaries must not
exceed the diffuse $\gamma$ background.

\section{A symmetric Universe}

\subsection{Sketch}

My colleagues Bel\'en Gavela and Andy Cohen (and I) \mbox{\cite{CDG}}
have studied a variety of theoretical scenarios for
a symmetric cosmology. Tenable
alternatives share many features. In particular, by the time
of nucleosynthesis, the Universe must consist of large matter or
antimatter regions of uniform density, separated by narrow
interstices that are not (or no longer are) domain walls. The subsequent evolution of
this early state and its confrontation with the observed Universe are
independent from its origin, and may be studied separately
\cite{CDGl}.

Our models of the generation of separate domains of matter and
antimatter (DMAs) are based on a Zen maxim: {\it If you find a fork on
the road... take it!} Thus may domains of cluster size or bigger, as they depart
from the horizon during inflation, take ``roads'' leading to
the two possible signs of the baryon-number excess.

Our DMA Universe is not unlike a
 magnetic material cooled below its Curie point. Even a flawless
material, if large enough and cooled rapidly, would not become a
domain with a single magnetic direction, for the speed at which the
information travels (the spin-wave velocity) is finite. Our analogue
of magnetization is the field the sign of whose phase determines
the baryon or antibaryon excess. Let
$\Delta\equiv n_B-n_{\bar B}$. The
difficulty for us is to end
up with domains of a very well defined
$\Delta\simeq \pm |\Delta_0|$ (only ``up'' or ``down'' magnetization,
but very little in between). We must accomplish this, because
if there were
domains with different baryon (or antibaryon) densities --and thus
different mass densities--  the abundances of primordial elements and
the CBR's temperature would show unacceptable inhomogeneities.
As it turns out, this constraint is not unduly difficult to satisfy.

Ours is an inflationary scenario and it shares with others the
necessity to choose an arbitrary set of parameters (there being no
theory of everything, in spite of millenarian claims to the
contrary). To the conventional lore we must add one arbitrary
parameter  that determines the average size $d_0$ of the DMAs. It turns
out not to be difficult to construct a theory for which the
distribution of DMA sizes is very sharply cutoff for sizes smaller than a given $d_0$.
This is good, for the occasional small antimatter domain in a larger
matter region would represent a serious observational problem.

As baryogenesis (and anti-baryogenesis) proceed as in conventional scenarios, 
we are left with
a patchwork Universe of regions of (current)
correlation-size $d_0$ randomly
containing only matter or only antimatter, separated by contact zones
of negligible width, relative to $d_0$. To figure out the fate of these
frontiers --as the Universe evolves from a dense plasma of many
species of particles to its present status-- is a lengthy but
edifying exercise in conventional physics.

One good reason not to indulge in a more detailed description
of our models is that they belong to the very general class that we
shall now proceed to infirm.

\subsection{Annihilation is inevitable}

From afar, the only way to tell about the presence 
of both matter and antimatter is to observe the direct or
indirect effects of annihilation. Imagine a scenario in which
matter and antimatter are separated by voids. How could it
possibly be refuted? As it turns out, the
observed uniformity of the CBR excludes the putative
voids, independently of whether they separate matter from matter
or matter from antimatter (topological walls could also do
the job, but their parameters would have to be very arbitrarily
concocted if they are to act as matter--antimatter buffers and
yet not contribute in excess to the universal
energy density).

At a temperature $\sim\!0.25$~eV, corresponding to
a redshift
$z_R\sim 1100$
the primordial plasma turned to neutral atoms
(recombination) and the radiation decoupled from ordinary matter
(last scattering).  The transition to transparency occurred during an interval $z_R\pm 100$, 
endowing the last scattering ``surface'' with a current (expanded)
width of $\sim\!15$~Mpc (an angular bracket $\Theta_{LS}\sim 8'$
in the transverse
direction). This angle is the
resolution of the ``picture'' of the CBR: smaller features
at recombination cannot be discerned.
(Some of the precise numbers I quote are specific to a 
ten--billion--year--old, critical, dark--matter--dominated Universe, 
but the
conclusions do not depend on this particular choice).

Voids would be non-homogeneities in the
baryon density. Such fluctuations are damped, at $z<z_{R}$,
by photons diffusing out of the
 over-dense  regions and dragging
matter along with them. By recombination, inhomogeneities with current
sizes $< 16$ Mpc would be destroyed by this mechanism \cite{Silk}. 
This bound fortuitously coincides with $\Theta_{LS}$:  
voids large enough to survive until recombination would have been
seen in the CBR. Since they are not, we conclude that matter and antimatter regions must be in contact after recombination. The {\it minimal} signatures of a 
baryon symmetric Universe
result from annihilations taking place  at $z<z_R$.

The evolution of a  nearly uniform primordial Universe into today's splotchy structure is not understood well
enough  to state its effect on matter--antimatter annihilation.
To avoid immediate trouble, it must be unlikely for a galaxy or a
cluster to contain comparable amounts of matter and antimatter. For
such structures to grow (rather than ``to ring'')
from the gravitational evolution of a mass
overdensity, a ``Jeans condition'' must be satisfied, the sound
travel time across the inhomogeneity $l/v_s$ must be longer than the
gravitational collapse or free fall time $1/\sqrt{G\, \rho}$. 
That is, $G\, l^2 \, \rho \ge v_s^2$.
If equality were approached for a region containing both matter and
antimatter, annihilation would reduce $\rho$, driving the system
away from collapse. 
Thus, one does not expect a matter--antimatter domain boundary
 to cut through a galaxy or cluster of galaxies. Grown-up density inhomogeneities should be of uniform composition.
To be conservative in assessing annihilation signatures
we must ``turn them off''  as soon as structure formation
becomes non-linear at some scale. For the corresponding redshift  we 
follow Peebles \cite{Peebles2} in adopting
the value $z_S\simeq 20$.  A 50\% up or down modification of this choice does
not affect our conclusions.

Annihilation is inevitable in the interval $z_R>z>z_S$.

\subsection{Explosive dynamics}

Imagine placing two semi--infinite gaseous regions,
one containing hydrogen, the other anti-hydrogen, in contact
along a plane.
 One's first impression is that the result would be a fairly violent run-away process, the dream of the combustion engineer: the gases would move towards a 
zone of overlap and annihilation, the annihilation products
would heat the gases, which would move faster towards annihilation,
producing more heat... 

For the matter densities characteristic of the early Universe, one's first impressions are not always right. For one thing,
 at redshifts $z<z_R$, the Universe is quite transparent to the photons resulting
from $p \bar{p}$ or $e^+ e^-$ annihilations. A small fraction of these photons interacts, but they deposit their (redshifted) energy far away from an annihilation region: they do not trigger an explosive reaction. 
Apart from irrelevant neutrinos, the only other stable ashes of H--$\overline{\rm H}$ annihilations are electrons and positrons of tens of MeV energies, made in $\pi \rightarrow \mu \rightarrow e$ decays.
 Their behaviour in the early Universe is peculiarly
complicated, and their role is crucial. In practice, the difference
between what electrons and positrons ``do'' is irrelevant, and I shall
refer to both species simply as electrons.

The electromagnetic shower made by the electrons is unlike
anything you have seen in the laboratory.
The electron energy loss along its trajectory, $dE/dx$, is dominated by scattering, not on the ambient matter, but
on the ambient light: the CBR. The CBR photons are
Compton scattered by the relativistic electrons
from their thermal energy to energies $\sim (E_e/m_e)^2$ times 
larger. The up-scattered photons have enormous ionizing
cross sections on hydrogen. As a result, at distances within 
the electron range from a layer of matter--antimatter contact,
and for the reckoned annihilation rate, 
the plasma stays fully ionized after recombination,
during all the epoch $z_R>z>z_S$ of interest to us. 

In the medium they keep ionized,
less than 1\% of the energy of the annihilation electrons is lost in 
collisions with ambient electrons or nuclei. But this small fraction is
 sufficient to produce, in the vicinity of a region where annihilations
 are taking place, a significant heating of the ambient matter, 
whose thermal history thus departs from the standard evolution. 

The thermalized energy deposited by electrons increases the
rate at which the matter and antimatter fluids converge towards
their annihilation, accelerating this process and the subsequent
local heating. While in the absence of this reheating it is possible
to give analytical approximations to the quantity of interest 
(the annihilation rate as a function of time), the 
real problem requires a numerical solution to the
equations for the fluids' motion. 

\subsection{Just a few equations}

Various approximations are 
 adequate to our analysis, at least in the interval
$z_R>z>z_S$.
All matter particles 
 maintain a common local temperature, not necessarily 
coinciding with the CBR temperature,
which follows its conventional redshift dependence 
$T_\gamma=T_0\, (1+z)$, $T_0\sim 2.7$ K. Elements
heavier than hydrogen can be neglected. 
The decay products of $e^+ e^-$ annihilation play an 
insignificant role, the process can be ignored.
Let $A(z)$ be the length scale over which matter and antimatter
fluids overlap
and annihilate, let $D(z)$ be the size of the domain
depleted by motion towards the annihilation region,
and let $L(z)$ be
the width of the region heated by the products of annihilation,
all as in Fig.~2.
At all times $L>D>A$ by one or two orders of
magnitude. The minimum 
size of a matter or antimatter domain evolves as $d=d_0/(1+z)$
and turns out to be larger than the other relevant scales.
Thus the curvature
of the boundary surfaces between domains can be
neglected: the fluid motion and annihilation problem
is one-dimensional,
with mirror symmetry between matter and antimatter. 

\begin{figure}
\epsfysize=2.8in
\centerline{\epsffile{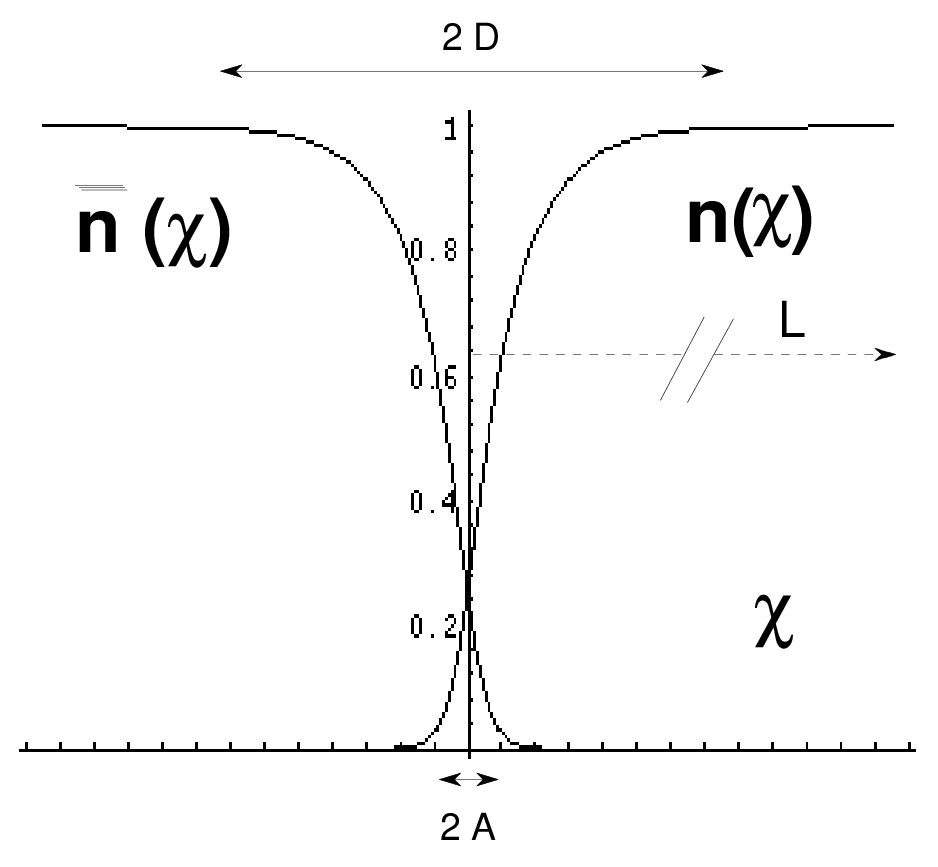}}
\caption{An artist's impression (not to scale)  
 of the density profile of matter
and antimatter at a domain boundary.}
\label{fig:contact}
\end{figure}



Let $R$ be the universal scale factor, and $\chi$ a comoving 
variable, with $\chi=0$ at the symmetry plane.
Let   $n$, $v$, $T$ and $P$ be the proton (or electron)
number density, velocity, temperature and partial pressure.
The antimatter quantities are $\bar n(\chi)=n(-\chi)$,
$\bar v(\chi)=-v(-\chi)$, etc. Baryon-number conservation
dictates:
\begin{equation}
  {\partial n \over \partial t} 
  + 3 \; {\dot{R} \over R}\;n
  +{1\over R}\; {\partial \, (n \, v) \over \partial \chi}=
- \, \langle \sigma_{\rm Ann}(p \bar{p})\,v({p \bar{p}}) \rangle \; n \,  \bar n \;.
\label{number}
\end{equation}
Energy conservation can be expressed as:
\begin{equation}
  {\partial P \over \partial t} 
  + 5 \; {\dot{R} \over R}\; P
  +{1\over R}\; v\; {\partial   P \over \partial \chi}
+{5 \over 3}\; {1\over R}\; P\; {\partial    v \over \partial \chi}=
 { 1 \over 2} \, n_\gamma \, \Gamma_{e \, \gamma} 
\, n \, \left( T_\gamma - T \right)
+{H \over 3}~. 
\label{pressure}
\end{equation}
The conventional first term on the r.h.s. describes the heat-bath effect
of the CBR, with $\Gamma_{e\gamma}$ the
rate of plasma--photon energy transfer: 
\begin{equation} \Gamma_{e\gamma}  = 
\thomp\; {\pi^2\over \zeta(3)}\; {T_\gamma \over m_e}  \end{equation}
and $\sigma_t$ the Thompson cross section.  In the second term on 
the r.h.s. of
Eq.~(\ref{pressure}), $H$ is the ``heat function'': 
 the energy
deposited per unit volume and time by the annihilation debris
in the ionized plasma. 
Let $dl=R\, d\chi$ and let $\langle dE/dl \rangle$ be the decremental
energy loss to the plasma by a single electron, in the direction $\vec l$
orthogonal  to the symmetry plane,
averaged over the electron emission angles relative to 
$\vec l$. The heat function is simply $H=J_e\, \langle dE/dl \rangle$,
with $J_e$ the current of annihilation electrons. The electron current
(into one side of the annihilation zone) is half of the $p \bar p$
annihilation current, $J_p$, times the electron multiplicity (roughly 3.8).
Finally, $J_p$ is the total annihilation rate
per unit surface orthogonal to $\vec l$:
\begin{equation} 
 J_p
=  \int \langle \sigma_{\rm Ann}(p \bar{p})\,v({p \bar{p}}) \rangle \;
n \,  \bar n \; dl ~.
\label{JeJp} 
\end{equation} 

Momentum conservation results in the third and last
of the fluid motion equations: 
\begin{equation}
  {\partial v \over \partial t} 
  + {\dot{R} \over R}\;v
  +{1\over R}\;v \, {\partial  v \over \partial \chi}
  +\,{1\over R}\; {1\over n}\; {1\over m_p}\; {\partial P  \over \partial \chi}
={ 1 \over 2} \, {m_e \over m_p}\, n_\gamma \, \Gamma_{e \, \gamma} \, v
 +  {H\over 2n\,m_p \, c}\;.
\label{velocity}
\end{equation}
The first term on the r.h.s. describes how the proper
motion of the fluid is damped  by friction of its
electrons against the CBR. 
The last term is the
momentum deposited in the reheating zone
by $e^\pm$ from $\bar{p} p$
annihilation. 

The solutions to Eqs.~(\ref{number})--(\ref{velocity})
depend on various cosmological parameters, notably
the baryon to photon ratio $\eta\equiv n_B/n_\gamma \simeq
n/n_\gamma$. We choose a value at the lower end of the
observationally allowed domain ($\eta=2\times 10^{-10}$) since
we are interested in the minimal annihilation signals.
Their dependence 
on the rest of the parameters (Hubble 
and cosmological constants, departure from closure)
within their empirically allowed range
does not amount to more than a factor $\sim 3$
and does not affect the conclusions.
I shall give results for $H_0=75$ km/s/Mpc, $\Lambda=0$,
$\Omega=1$, implying $R\propto t^{2/3}$ in the
redshift interval of interest.

The solutions to Eqs.~(\ref{number})--(\ref{velocity}) are qualitatively
different at large and small redshifts.  For $z>400$ the CBR drag force dominates so
that the motion is diffusive. 
In these
early times the CBR is also an effective heat bath that keeps matter in
thermal equilibrium with radiation everywhere, even in regions reheated by annihilation. This early diffusive period has a welcome
consequence: all memory of the initial conditions is lost as the
fluid evolves. The post-recombination annihilation signal does not depend
on the (pre-recombination) time at which matter and antimatter 
domains first come into
contact. For $z<400$ the
pressure-gradient dominates the CBR-drag so that the
fluid motion is ``hydrodynamic''. Moreover,
heating due to the annihilation electrons plays
an important role. Positive feedback sets in to increase the
annihilation current. This
potentially runaway process is eventually quenched by rarefaction. For
$z<30$, the Universe is so sparsely populated that cooling by expansion
dominates the evolution of the matter temperature.

The terms representing CBR drag and annihilative heating  ensure that the matter
 temperature in the moving fluids
is spatially constant. 
 The computed value of this  temperature is shown  as a
function of $y=1+z$ in Fig.~3, where it is compared with the 
temperature
beyond the reheating zone (as it would be in an all-matter Universe).

\begin{figure}
\epsfysize=3.5in
\centerline{\epsffile{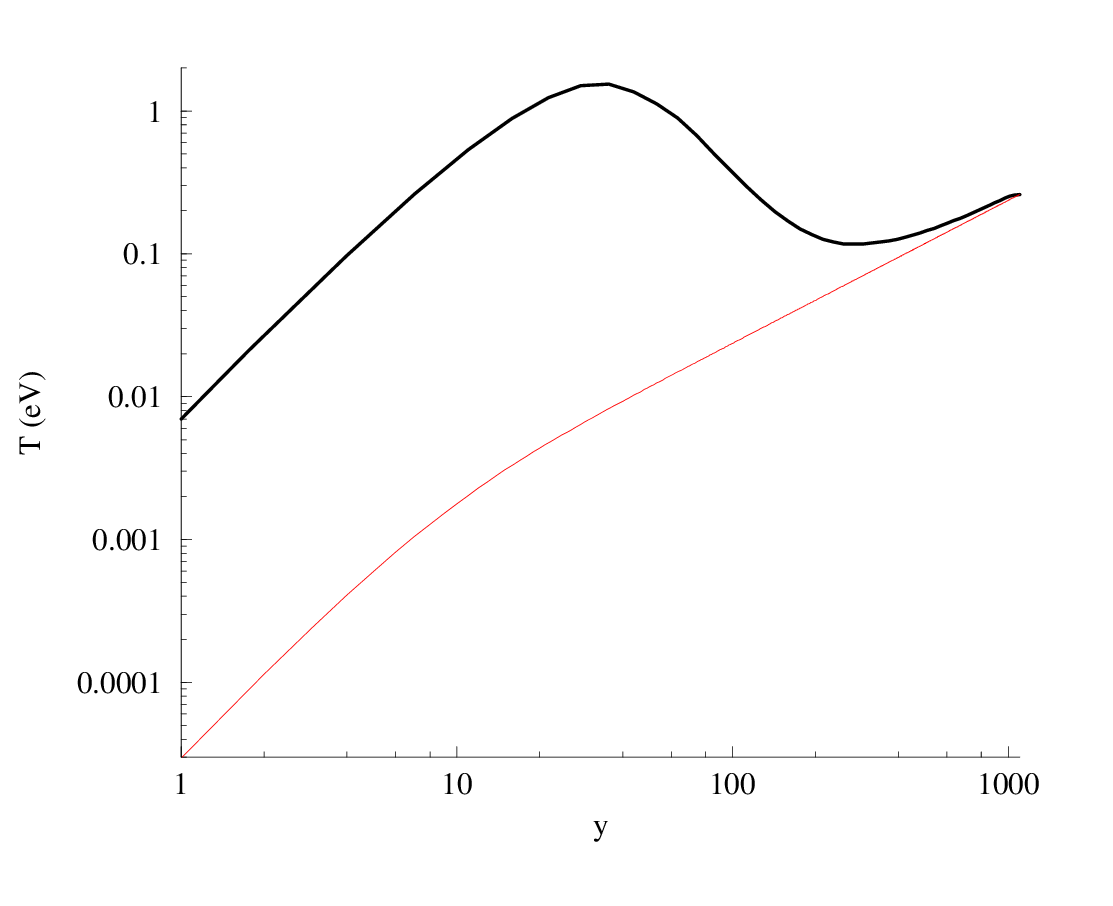}}
\caption{Temperature (in eV) as a function of redshift
$y=1+z$. The upper curve is our numerical solution.  In the lower
(conventional) curve, reheating is ignored$^1$.}
\label{fig:temperature}
\end{figure}



\section{Annihilation signatures}

The solution of the fluid annihilation equations
provides the annihilation current $J_p$ of Eq.~(\ref{JeJp})
as a function of redshift. We do not live close to
an annihilation zone, and we are mainly interested
in the non-local, or ``diffuse''
collective effects of annihilation throughout the
Universe. These depend on the
rate of annihilation per unit {\it volume\/} and time (averaged over a
region large enough to encompass various matter and antimatter domains) which is
$J_p/d$, with $1/d$ their average surface-to-volume ratio ($d$
expands as $d=d_0/y$).  

\subsection{The cosmic diffuse gamma-ray background (or CDG)}

Let $\Phi(E)$ denote the (lab) inclusive photon
spectrum
per $p \bar{p}$ annihilation. The average number of photons made per unit volume, time and
energy is $\Phi(E)\,{J_p/ d}$. 
Let $N(E,y)$ be the flux of annihilation photons (per unit time, area and
solid angle) reaching a point with redshift parameter
$y=1+z$. This flux
evolves according to a ``renormalization group'' equation:
\begin{equation}
  \left( y\,{\partial\over \partial y}
    +E\, {\partial\over \partial E} -2 \right)\,N(E,y) =
 \int g(E,E',y)\,N(E',y)\,dE'
-{c\; \Phi(E)\over H_0\, y^{3/2}}\, {J_p(y)\over 4\, \pi \, d(y)} 
\label{transport}
\end{equation}
where the first term on the r.h.s. is a correction for 
photon rescattering  and the second is the
annihilation source.  The current flux is $N(E,1)$.
In solving this equation, we ``switch on'' the
source only in the redshift interval $z_R>z>z_S$.

The observed CDG flux is compared to our (conservative
lower limit) flux $N(E,1)$
in Fig.~4.  In
the 2--10 MeV energy range, recent preliminary COMPTEL
satellite measurements
\cite{comptel} lie roughly an order of magnitude below the earlier
balloon data \cite{gam1}$^-$\cite{gam6}.
Two theoretical results are shown: the upper curve 
is for $d_0=20$ Mpc, the lower one for $d_0=1000$ Mpc.
The $N(E,1)$ spectrum is redshifted from the 
spectrum at production (which peaks at $E\sim 70$~MeV), and is slightly
depleted at the lowest energies  by attenuation.
  The $d_0=10^3$~Mpc result is barely compatible with
the balloon data, and an order of magnitude above the satellite data.
To reach agreement, $d_0$ must be comparable to or larger than the current
horizon.

\begin{figure}
\epsfysize=3.5in
\centerline{\epsffile{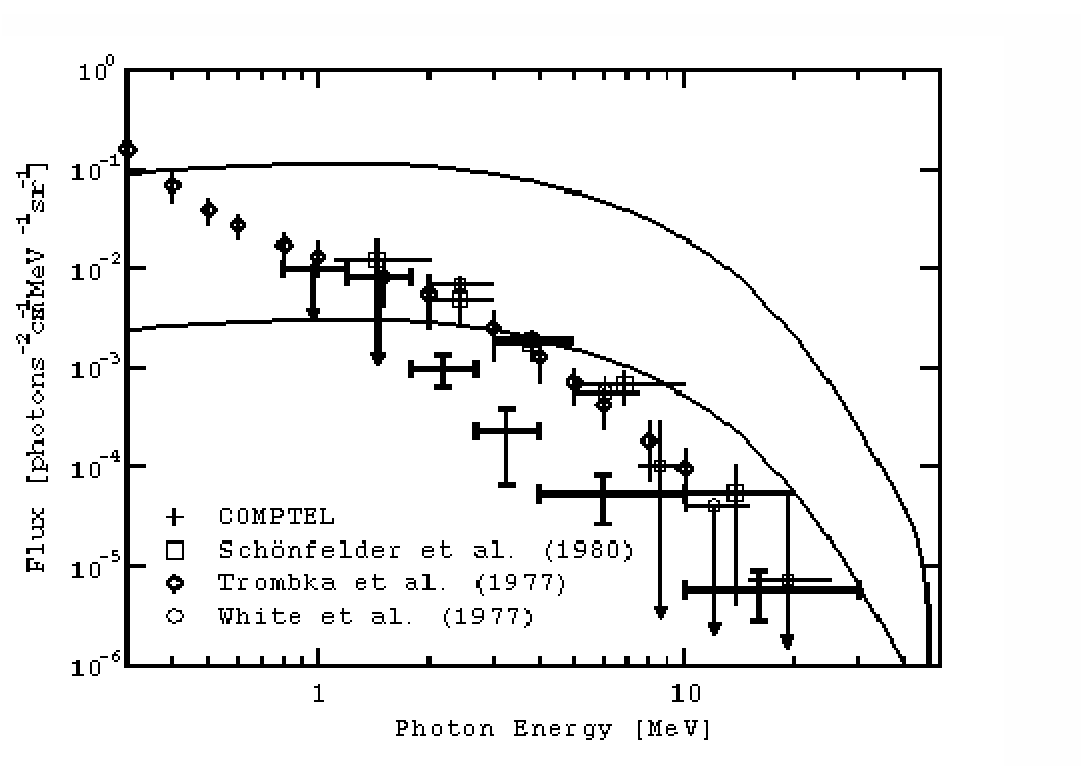}}
\caption{Data and expectations for the number-flux diffuse
 $\gamma$-ray spectrum$^1$.}
\label{fig:gammas3}
\end{figure}



The conclusion is clear: the diffuse gamma-ray
observations completely exclude a Universe containing significant
amounts of antimatter.  Are the constraints from the much-better measured CBR
comparably stringent?

\subsection{Distortion of the CBR}

Matter--antimatter annihilation would make the CBR spectrum deviate
from its thermal distribution.  A flux of
``Comptonized'' photons is produced as the
annihilation electrons scatter on the CBR.
 These populate a UV region of energies
wherein the hydrogen photoionization cross section is extremely large.
The UV photons reaching the border of the
ionized domain deposit
their energy in photoionization reactions. Through these secondary
interactions, a fraction $f(y)$ of the energy of the original
annihilation-product electron ends up as heat.
(The fraction $f(y)$ is very weakly dependent  
on the initial electron energy, and is always
close to unity;
a lengthy exercise demostrates that
 $f(y)$ rises from $\sim 1/2$ at $y=10$ to saturate
close to  100\% above $y=300$.) The heated ambient
electrons, in a third step, distort the CBR.

The Sunyaev--Zel'dovitch parameter $Y$
characterizes the thermal distorsion
as a frequency-dependent 
``temperature'', a function of $x=\nu/T_0$:
\begin{equation}
T(x)\sim T_0\,\left[1+Y\,\left({x\,(e^x+1)
\over e^x-1}-4\right)\right]~.
\end{equation}
The COBE results translate into $|Y|<1.5 \times 10^{-5}$.

On average, and per $p \bar p$ annihilation, some 
$\Delta E\sim 320$ MeV of energy
are carried away by electrons. The
universally averaged energy per unit volume and 
redshift interval deposited
by the photoionizing interactions is:
\begin{equation}
\Bigl | {d\epsilon\over dy} \Bigr | ={1\over H_0\, y^{5/2}}\,
{J_p(y)\,\Delta E\over d(y)}\, f(y)
\label{dEdy}~.
\end{equation} 
Subject to this heat, the ambient electrons interact with
the CBR, resulting in a 
predicted:
\begin{equation}
Y\simeq {15\over 4\,\pi^2}\,\int_{y_{S}}^{y_{R}}\,
dy\;{1\over T_\gamma^4(y)}\;{d\epsilon\over dy}
\label{Yvalue}
\end{equation}
where $y_{S}$ and $y_{R}$ are our consuetudinary redshift cut-offs.

To compute the value of $Y$ in Eq.~(\ref{Yvalue}) one has to use the
current $J_p$ from the solution to our fluid 
equations.  For $d_0=d_{\rm min}=20$ Mpc
the result is
$Y=4.6\times 10^{-4}$, exceeding by over
one order of magnitude the COBE
limit. To have theory comply with this observational
stricture we must have $d_0>700$ Mpc.
This limit is  stronger than the one stemming
from X-ray emitting clusters, but it is not as strong
as the one we obtained from the diffuse $\gamma$-ray background.

\section{Caveats?}

We have tried to find a weakness in our {\it no-go theorem}
stating that the Universe is indeed asymmetric in its 
matter/antimatter constituency. ``Isocurvature voids'' and primordial 
magnetic fields
of a very specific nature are the only caveats 
we have found that we cannot
exclude on the basis of observations and well established physics. 

By isocurvature voids, I mean a scenario in which matter and antimatter islands
would be separated by interstices with vanishing baryon density, but the photon
 distribution would be
uniform.  In models with isocurvature fluctuations, 
disfavored by observations of the CMB  and of
galaxy clustering \cite{Ue}, our
arguments about matter and antimatter necessarily touching at recombination
would not apply.
We have not pursued this line of thought any further.

The effect of magnetic fields that are sufficiently strong
and disordered (small correlation length)
would be to shorten significantly  the distance over which
annihilation electrons deposit their energy. The reheating due
to these electrons becomes more efficient, the effect 
goes in the direction of increasing the annihilation rate and
improving our bounds. This is unless the electron reach
becomes so short that the nature of the solution to our
fluid motion equations changes drastically.

There is no known way to
generate magnetic fields in the pre-recombination
plasma by conventional dynamo effects;  their
production from the latent heat of some first-order
phase transition is the most often invoked 
hypothesis \cite{nordics}.
If the primordial Universe was ever at a temperature
exceeding the mass of the weak vector bosons
($T>T_W\sim 100$ GeV) it is natural to assert that 
electromagnetism was ``born'' in the phase transition
that possibly occurred as the electroweak symmetry
``broke''. The question of whether
such a transition could generate magnetic fields is debated, 
the details of the resulting field strength and structure
are a matter of guesswork. Typical assumptions are a field 
energy density comparable to that of the other $\sim 100$
degrees of freedom acting at that time
(or $B\sim 10^{24}$ gauss) and a correlation length
one to three orders of magnitude smaller than the horizon
(a mere 1.4 cm by then).

We have studied the evolution of primordial fields
originating in an electroweak or QCD transition
and we find that their effect is to increase
the annihilation rate and strengthen our
conclusions. But we cannot entirely 
exclude the existence of an {\it ad hoc} magnetic
field structure with a correlation length much shorter than
the ones we have studied, nor the possibility
that the current understanding of magneto-hydrodynamics be insufficient to reach
a definite conclusion.

\section{Conclusions at 3/4 of the way}

By the summer of 1974, a group at Brookhaven had detected,
in the debris of hadronic collisions, a narrow
peak at 3.1 GeV in the invariant mass of $e^+ e^-$ pairs.
Practically every theorist would (then)
have said that one could prove on general grounds,
and beyond the shadow of a doubt, that something made with
a hadronic-sized cross section and weighing as much as
3.1 GeV, HAD to be very broad: the data HAD to be wrong.
 This is to say that (though the standard
model has accustomed us otherwise) theorists are apt
to miss a point, if the point is big enough.

Astrophysics is a subject wherein surprises also pullulate.
Quasars, pulsars, invisible mass, gamma-ray bursts and high
energy cosmic rays would have been difficult to guess.

Finally, nothing can compete with {\it direct} limits or observations, 
e.g. of antinuclei in the cosmic rays. The above are three reasons to 
look for these alien creatures.  I proceed to discuss one of the efforts in this direction.

\section{Extragalactic cosmic rays}
To reach
us, a cosmic ray from a more distant galaxy (or antigalaxy) must have
been able to exit from it, to travel all the way here, and to
penetrate the galactic disk.  This is feasible, as I proceed to
 outline (for details and numerous references, see Ref. 24).
Intergalactic travel is the least problem.
 From the time of galaxy formation, the density of intergalactic
matter has been far too small to intercept travelling nuclei. There
are no solid grounds to believe that intergalactic space is permeated by
magnetic fields strong enough to encumber the straight voyage of an
energetic charged particle. A relativistic particle could
reach us from the confines of the visible Universe.

Our galaxy has a microgauss magnetic field of complex
structure. The average (charged) cosmic ray meanders around the
galaxy for a ``confinement time'' at least a thousand times longer
than the few thousand years it would take it to cross the galactic
disk forthright. Could cosmic rays never escape, as in ``closed
galaxy models''?

Our knowledge of the history of cosmic rays is based on the study of
their chemical and isotopic composition. The relative abundance of
the various elements at the location where the rays are accelerated
is presumed to be akin to that of the solar system. The ``arrival''
abundances are indeed generally similar to the solar ones, with a
pronounced odd--even $Z$-variation and a peak for Fe, as befits
elements that have been made in stars (H and He are primordial,
and underabundant in cosmic rays).

Certain overabundances of cosmic rays
(Li, Be, B and Sc to
Mn)  are attributed to the fragmentation
of larger primary nuclei. The ``leaky box'' model, wherein cosmic
rays are magnetically confined but have a chance of escaping the
galaxy, is the simplest one to fit these observations. The
(excellent) fit results in a value for the mean traversed column
density (roughly 10 g/cm$^2$, for a kinetic energy of 1 GeV
per nucleon). The confinement time is more directly bracketed (to
$25 \pm 10$ Myr) by the abundances of $^{10}$Be, whose lifetime is
2.3 Myr, and $^9$Be, which is stable. Closed models are definitely
excluded \cite{closed} by their inability to explain the $^3$He-to-$^4$He ratio.

Cosmic rays are obstructed not only by the Earth's magnetic field,
but also by the outflowing solar wind. Similarly, the ``galactic
accessibility'' --the probability that an alien ray penetrates our
galaxy-- is affected by magnetism and by the halo galactic wind,
driven by supernova explosions. The estimate
 \mbox{\cite{APST}} is that, at a kinetic energy/nucleon of 1~GeV, the
 accessibility may be 10--50\%.

Finally, one must estimate the fraction $E/G$ of extragalactic to
galactic rays.
Meteorite observations demonstrate that the cosmic-ray flux has been
constant for at least 4 Gyr, consistent with an equilibrium between
production and leakage. Assume the $t_{\rm res}\sim\! 25$ Myr
cosmic-ray residence time in our galaxy to be typical. Galaxies have
been around for some $t_{\rm gal}\sim\! 15$ Gyr. The volume fraction
currently occupied by galaxies is ${\rm f}\sim\! 10^{-7}$. Clearly,
in a steady state of cosmic-ray production, moderate absorption, and
subsequent leakage, $E/G\sim\! {\rm f}\,t_{\rm gal}/t_{\rm res}\sim\!
6\times 10^{-5}$, to be further reduced by the accessibility factor.

All of the above considerations enter the calculation of the
$\overline{\rm {He}}$/He fraction displayed in Fig.~5, for a hypothetical
Universe made of equally many DMAs of size $d_0=20$ Mpc. The fraction
is probably an underestimate; it assumes that cosmic rays diffuse in
a maximally disordered magnetic field sustained by a hot intercluster
plasma, once argued to be required to explain diffuse X-rays
\mbox{\cite{Plasma}} and now excluded \mbox{\cite{noplasma}}. The
difference between diffusive and straight travel  reduces the
effective reach of extragalactic cosmic rays from $l\sim 3000$ Mpc to
$l\sim 150$~Mpc.

\begin{figure}
\epsfysize=3.0in
\centerline{\epsffile{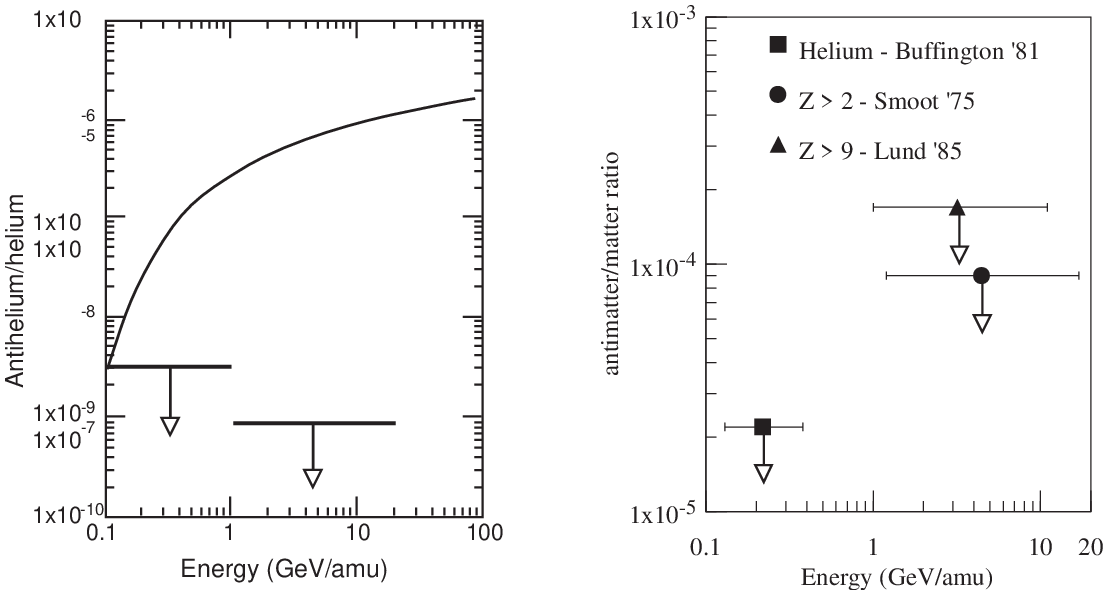}}
\caption{Left:  $\overline{\rm {He}}$/He ratio, expectation in a symmetric
 Universe and reach of AMS (plateaus).  Right:  current limits on antinuclei in
 cosmic rays.}
\label{fig:gammas}
\end{figure}



\section{The Alpha Matter Spectrometer}


Impervious to the relaxed progress of the theoretical work I have
described, a team whose spokesperson is Sam Ting
\mbox{\cite{AMS,AMSproposal}} has been busy designing and
constructing the {\it Alpha Matter Spectrometer} (AMS), a device to
be flown in Earth's orbit, meant to improve by many orders of
magnitude our knowledge of cosmic rays.

The main detector of AMS is a small ($\sim\!
1$ m$^3$) charged-particle spectrometer measuring charge (squared)
from the energy deposition along a track, velocity with a
time-of-flight device, and momentum (over charge) with a
tracker surrounded by a magnet.
The technological AMS break-through is in its magnet. Unlike that of
previous projects \mbox{\cite{Astromag}}, it is not a superconducting
magnet necessitating liquid-He refrigeration, an added nuisance. It is a permanent magnet made
of  Nd--Fe--B crystals that can sustain an unprecedentedly strong
magnetic field.
Some of the current limits on cosmic--ray antinuclei,
as well as the expected reach of AMS 
(a three to four orders of magnitude leap into unexplored land) are shown in Fig. 5.  

The AMS proposal was approved by NASA and the DOE in  1995, and
scheduled for a first test flight in the space shuttle, to be
launched on the 2nd of April of 1998. The significance of this date
should be clear: it is the anniversary of Hans Christian Andersen,
the master of fairy tales. This maiden flight should last two weeks,
the AMS detector will sit in the shuttle's cargo bay, which can be
opened up as a particularly posh convertible. The plan is to fly
part-time with the shuttle ``upside down'', its opening facing the
Earth, to measure directly the ``albedo'' of cosmic rays bouncing up
after hitting the top layers of the atmosphere. If all goes well, the AMS
would be added, in the year 2000, to the International Space Station
{\it Alpha} for continued operation, lasting several years.
My hunch is that AMS was not approved because of its capability to
search for cosmic-ray antinuclei. That is far too long a shot to move
a committee. It must have been approved because of the observational
improvements it will bring in ``orthodox'' areas of physics such as
the measurement of the low-energy $\bar{p}$ flux, an indirect window
into the annihilation of halo 
super-wimps \cite{Gordy}, as we have already
mentioned.

Even more conservatively, AMS will constitute an enormous step in our
knowledge of the spectra of conventional (matter) cosmic-ray nuclei.
Various flux ratios, such as ${^9}$Be/$^{10}$Be, B/C, ${^3}$He/${^4}$He will be measured with unprecedented precision, and we have
discussed how crucial they are to our understanding of cosmic rays.
Most
importantly, AMS will be the first detector sensitive to extragalactic cosmic
 rays, and this is where the serendipitous surprises may lie.

\section*{Acknowledgments}
I am indebted to Bel\'en Gavela for innumerable debates and a critical reading
of the manuscript.  I am also indebted to many members of the AMS
collaboration, in particular Steve Ahlen, for many discussions.  I have learned
on a lot of subjects from too many people to quote, but I shall make 
exceptions for Sid Redner on reaction kinetics and Juan Garcia Bellido on inflation.

  \end{document}